# Stochastic thermodynamics for delayed Langevin systems

Huijun Jiang, Tiejun Xiao, Zhonghuai Hou[1]

*Hefei National Lab for Physical Science at Microscale & Department of Chemical Physics*

*University of Science and Technology of China, Hefei, Anhui, 230026, P.R.China*

**Abstract:** Stochastic thermodynamics (ST) for delayed Langevin systems are discussed. By using the general principles of ST, the first-law-like energy balance and trajectory-dependent entropy $s(t)$ can be well-defined in a similar way as that in a system without delay. Since the presence of time delay brings an additional entropy flux into the system, the conventional second law $\langle \Delta s_{tot} \rangle \geq 0$ no longer holds true, where $\Delta s_{tot}$ denotes the total entropy change along a stochastic path and $\langle \cdot \rangle$ stands for average over the path ensemble. With the help of a Fokker-Planck description, we introduce a delay-averaged trajectory-dependent dissipation functional $\eta[\chi(t)]$ which involves the work done by a delay-averaged force $\bar{F}(x,t)$ along the path $\chi(t)$ and equals to the medium entropy change $\Delta s_m[x(t)]$ in the absence of delay. We show that the total dissipation functional $R = \Delta s + \eta$, where $\Delta s$ denotes the system entropy change along a path, obeys $\langle R \rangle \geq 0$, which could be viewed as the second law in the delayed system. In addition, the integral fluctuation theorem $\langle e^{-R} \rangle = 1$ also holds true. We apply these concepts to a linear Langevin system with time delay and periodic external force. Numerical results demonstrate that the total entropy change $\langle \Delta s_{tot} \rangle$ could indeed be negative when the delay feedback is positive. By using an inversing-mapping approach, we are able to obtain the delay-averaged force $\bar{F}(x,t)$ from the stationary distribution and then calculate the functional $R$ as well as its distribution. The second law $\langle R \rangle \geq 0$ and the fluctuation theorem are successfully validated.

---

[1] To whom correspondence should be addressed. Email address: hzhlj@ustc.edu.cn.

## 1. INTRODUCTION

Nonequilibrium thermodynamics in small systems has become nowadays an active field in statistic physics, which has widely applications in nano and life systems [1,2]. As a result of the small system size, fluctuations are significant in small systems, such that the thermodynamic quantities become stochastic variables. This observation brings the attempt to establish nonequilibrium thermodynamics in the stochastic level, which is known as stochastic thermodynamics (ST) [3-5]. In an overdamped Langevin system, Seikimoto has considered the internal energy, work and heat, in order to interpret the first law along a stochastic trajectory [3]. Seifert goes one step further to define trajectory dependent entropy and stochastic entropy production [4,5], where the total entropy change $\Delta s_{tot}$ along a path, which is the summation of the change of system entropy $\Delta s$ and of medium entropy $\Delta s_m$, satisfies the remarkable fluctuation theorems (FT) [6-11]. ST has been successfully applied to the two level optical system [12], forced Brownian particles [13], mesoscopic chemical reaction network [14], and state transition processes in bimolecules [15], and so on.

On the other hand, for the recent decades there has been considerable interest in delayed systems [16-26], whose dynamics are determined by both the present state $x \equiv x(t)$ and the past state $x_\tau \equiv x(t-\tau)$, where $\tau > 0$ is the delay time. In real systems, delay is usually ascribed to finite speed of transmission of matter or information, or some kind of feedback control. It has been shown that delayed systems may exhibit complex dynamic behaviors, such as delay-induced excitability [16], delay-induced oscillation [17], to list just a few. Delayed models have been widely applied to describe chemical kinetics [18-20], neural networks [21], physiological systems [22], optical devices [23, 24], population dynamics [25], economic systems [26], and so on. The presence of time delay offers strong non-Markovian property of the system, which leads to many open problems, letting aside their applications in many real systems, as mentioned above. Specifically, to the best of our knowledge, the energy balance and entropy production have been investigated via a response function method in a linear stochastic system with delay [27]. However, the ST for general delayed systems, especially proper interpretations of the second law and FT, have not been well addressed.

In the present work, we have considered the ST of stochastic systems described by delayed Langevin equations. By using the stochastic energetics approach [3] proposed by Sekimoto, a first-law-like energy balance can be established, wherein all the energy, work and heat dissipation can be functions of both $x$ and $x_\tau$. According to Seifert, one can define the trajectory dependent entropy based on the probability distribution function $p(x;t)$ for $x$ at time $t$. Entropy balance equation can be obtained with the help of a Fokker-Planck equation (FPE) which involves a joint probability density $p(x_\tau, t-\tau; x, t)$ that the state variable takes value $x$ at time $t$ and $x_\tau$ at time $t-\tau$. This defines a delay-averaged dissipation functional $\eta$, satisfying $\langle \Delta s + \eta \rangle \geq 0$, which can be viewed as the second law. In the absence of time delay, i.e., $\tau = 0$, $\eta$ equals to $\Delta s_m$, such that the conventional second law $\langle \Delta s_{tot} \rangle \geq 0$ holds. For $\tau > 0$, however, $\eta$ is different from $\Delta s_m$, and hence $\langle \Delta s_{tot} \rangle$ could be negative. The discrepancy between $\eta$ and $\Delta s_m$ can be intuitively viewed as a type of injected entropy introduced by the delay feedback. By using path integral for the FPE, we show that the integral FT holds for $R \equiv \Delta s + \eta$, i.e., $\langle e^{-R} \rangle = 1$. In the stationary state, a detailed FT $p(R)/p(-R) = e^R$ also holds for $R$. We apply these general results to a stochastic linear system with delay and external periodic force. With the help of inversion of the stationary distribution function, we are able to calculate the total dissipation $R$ as well as its distribution. The second law and FTs are successfully validated by numerical calculations.

## 2. STOCHASTIC THERMODYNAMICS

### 2.1. The First Law

Consider a stochastic system described by the following Langevin equation

$$\dot{x} = F(x, x_\tau; \lambda) + \xi(t), \qquad (1)$$

where $F(x, x_\tau; \lambda)$ is a systematic force and $\xi(t)$ thermal noise with correlation $\langle \xi(t)\xi(t') \rangle = 2D\delta(t-t')$ where $D$ is the noise intensity. In equilibrium, $D$ is related to the temperature $T$ by the Einstein relation $D = k_B T$ where the Boltzmann constant $k_B$ is set to unity in the present study. Generally, the force $F$ could arise from a conservative potential $V(x;\lambda)$ and/or a directly applied force $f(x, x_\tau; \lambda)$ as

$$F(x, x_\tau; \lambda) = -\partial_x V(x; \lambda) + f(x, x_\tau; \lambda). \qquad (2)$$

Here both terms may be time-dependent through an external control parameter $\lambda(t)$ varied according to some experimental protocol from $\lambda(t_0) \equiv \lambda_0$ to $\lambda(t_1) \equiv \lambda_{t_1}$. According to the stochastic energetics approach [3], the work increment applied to the system

$$dw = (\partial V/\partial \lambda) \dot{\lambda}\, dt + f\, dx \quad (3)$$

consists of contributions from changing the potential and that from applying the non-conservative force. Physically, one can identify the work done by the random force and the frictional force as the heat dissipation into the medium

$$dq = (\dot{x} - \xi(t))\, dx = F\, dx. \quad (4)$$

Multiplying both sides of Eq.(1) by $dx$ and then using Eqs. (3) and (4), one obtains

$$dV = dw - dq, \quad (5)$$

which can be viewed as the first-law-like energy balance equation corresponding to Eq.(1). Note that all these interpretations are similar to those for a system without delay, except that both the work and heat are now delay-dependent stochastic variables.

Now consider a stochastic path $\chi(t) = \{x(t)|_{t=t_0}^{t=t_1}\}$ that starts from $\phi_0 \equiv \{x(t)|_{t=-\tau}^{0}\}$ and ends at $x_{t_1}$ generated from Eq.(1). Integrated along $\chi(t)$, one reaches the following expression,

$$q[\chi(t)] = w[\chi(t)] + V(x_{t_1}, \lambda_{t_1}) - V(x_0, \lambda_0) \quad (6)$$

where

$$q[\chi(t)] = \int_{t_0}^{t_1} F\, \dot{x}\, dt \quad (7)$$

and

$$w[\chi(t)] = \int_{t_0}^{t_1} \left((\partial V/\partial \lambda)\dot{\lambda} + f\, \dot{x}\right) dt \quad (8)$$

are trajectory dependent heat and work, respectively.

## 2.2. The Second Law

As suggested by Seifert [4, 5], we can define as a trajectory dependent entropy of the system the quantity

$$s(t) = -\ln p(x(t); t) \quad (9)$$

where $p(x; t)$ is the probability for the state variable to take value $x$ at time $t$ no matter what value

$x_\tau$ takes, and $s(t)$ is evaluated along the stochastic trajectory $\chi(t)$. However, for a delayed stochastic system, it is not trivial to obtain $p(x;t)$. Very recently, T. D. Frank et al. suggested that $p(x;t)$ obeys the following Fokker-Planck equation (FPE) [28],

$$\partial_t p(x,t) = -\partial_x \left[ \int dx_\tau F(x,x_\tau) p(x_\tau, t-\tau; x,t) \right] + D\partial_x^2 p(x,t) \equiv -\partial_x j(x,t), \quad (10)$$

where $p(x_\tau, t-\tau; x, t)$ denotes a joint probability density that the state variable takes value $x$ at time $t$ and $x_\tau$ at time $t-\tau$, and

$$j(x,t) = \int dx_\tau F(x,x_\tau) p(x_\tau, t-\tau; x, t) - D\partial_x p(x,t)$$
$$\equiv \bar{F}(x,t) p(x,t) - D\partial_x p(x,t) \quad (11)$$

is the probability flux density. Herein, the quantity

$$\bar{F}(x,t) = \int dx_\tau F(x,x_\tau) p(x_\tau, t-\tau | x, t) \quad (12)$$

denotes a delay-averaged force (DAF) which does not depend on $x_\tau$, where $p(x_\tau, t-\tau | x, t) = p(x_\tau, t-\tau; x, t) / p(x,t)$ is the conditional probability that the system history takes $x_\tau$ at time $t-\tau$ given that the current state at time $t$ is $x$. The FPE can then be rewritten as

$$\partial_t p(x,t) = -\partial_x \left[ \bar{F}(x,t) p(x,t) \right] + D\partial_x^2 p(x,t) \quad (13)$$

Thanks to the FPE description, the change rate of system entropy is given by [4]

$$\dot{s}(t) = \frac{-\partial_t p(x,t)}{p(x,t)} - \frac{\partial_x p(x,t)}{p(x,t)} \bigg|_{x(t)} \dot{x}$$
$$= \left[ \frac{-\partial_t p(x,t)}{p(x,t)} + \frac{j(x,t)}{Dp(x,t)} \bigg|_{x(t)} \dot{x} \right] - \bar{F}(x,t) \dot{x} \quad (14)$$

where Eq.(11) has been used to get the second equality. By the same reasoning as in Ref.[4], one readily obtains that

$$\langle \dot{s} + \bar{F} \dot{x} \rangle = \left\langle \frac{j(x,t)}{Dp(x,t)} \bigg|_{x(t)} \dot{x} \right\rangle = \int dx \frac{j(x,t)^2}{Dp(x,t)} \geq 0, \quad (15)$$

where $\langle \cdot \rangle$ stands for average over the path ensemble, and $\int dx \partial_t p(x,t) = 0$ and $<\dot{x}|x,t> = j(x,t)/p(x,t)$ have been respectively used in the first and second equality. Integrated along a given trajectory $\chi(t)$, we have the following expression

$$\langle R[\chi(t)] \rangle \equiv \langle \Delta s[\chi(t)] \rangle + \langle \eta[\chi(t)] \rangle \geq 0, \quad (16)$$

where $\Delta s[x(t)] = \ln \dfrac{p(x_{t_1};t_1)}{p(x_{t_0};t_0)}$ is the system entropy change and

$$\eta[\chi(t)] = \int_{t_0}^{t_1} \overline{F} \, \dot{x} \, dt \quad (17)$$

defines a type of delay-averaged path-dependent dissipation functional.

One notes here that the quantity $\eta[x(t)]$ is different from the heat dissipation $q[x(t)]$ introduced in Eq.(7) because $\overline{F}(x,t)$ and $F(x,x_\tau,t)$ are generally not identical. The heat dissipation into the environment can always be identified with an increase of the medium entropy, no matter delay is present or not, i.e.,

$$\Delta s_m[x(t)] = q[x(t)]/T. \quad (18)$$

In the absence of delay, $\overline{F} = F$ and hence $\eta = \Delta s_m$, therefore Eq.(16) recovers the conventional second law $\langle R|_{\tau=0} \rangle = \langle \Delta s_{tot} \rangle = \langle \Delta s + \Delta s_m \rangle \geq 0$. However, in the presence of delay, the quantity $R$ is different from $\Delta s_{tot}$. Therefore, Eq.(16) tells that it is the functional $R[\chi(t)]$ rather than $\Delta s_{tot}[\chi(t)]$ that should be non-negative in a delayed stochastic system, which can be viewed as the generalized second law.

In other words, in the presence of delay, there is no rule to guarantee that the total entropy change to be non-negative. Intuitively, in a delayed system, one needs to apply the information at $t-\tau$ to the current state, thus introduce a type of entropy flux into the system. The agent who performs the delay-feedback works as a kind of 'demon'. Generally, the delay-induced entropy flux can be positive or negative. However, the system and the 'agent' as a whole, should obey the second law definitely. Our analysis shows that the total dissipation functional $R[\chi(t)]$ plays the key role.

**2.3. FTs for the Total Dissipation**

Given the validity of the FPE where the effect of delay has been averaged, Eq.(13), one may write down the probability of a given trajectory starting from $x_0$ as follows [29]

$$p[\chi(t)|x_0] = \exp\left(-\int_{t_0}^{t_1} dt\left[(\dot{x}-\bar{F}(x,\lambda))^2/4D + \partial_x \bar{F}(x,\lambda)/2\right]\right). \quad (19)$$

Similarly, the conditional weight $p(\tilde{\chi}(t)|\tilde{x}_0)$ of the time-reversed trajectory $\tilde{\chi}(t) = \{\tilde{x}(t)|_{t_0}^{t_1}\} = \{x(t_1-t)|_{t_0}^{t_1}\}$ starting from $\tilde{x}_0 = \tilde{x}(t_0) = x(t_1)$ under the time-reversed protocol $\tilde{\lambda}(t) = \lambda(t_1-t)$ is given by,

$$\begin{aligned}p[\tilde{\chi}(t)|\tilde{x}_0] &= \exp\left(-\int_{t_0}^{t_1} dt\left[(\dot{\tilde{x}}-\bar{F}(\tilde{x},\tilde{\lambda}))^2/4D + \partial_x \bar{F}(\tilde{x},\tilde{\lambda})/2\right]\right) \\ &= \exp\left(-\int_{t_0}^{t_1} dt\left[(\dot{x}+\bar{F}(x,\lambda))^2/4D + \partial_x \bar{F}(x,\lambda)/2\right]\right)\end{aligned} \quad (20)$$

Combine Eqs.(19) and (20), one can obtain

$$\eta[\chi(t)] = \int_{t_0}^{t_1} \bar{F}(x,\lambda)\dot{x}dt = \ln\frac{p(\chi(t)|x_0)}{p(\tilde{\chi}(t)|\tilde{x}_0)} \quad (21)$$

which builds the connection between the dissipation functional $\eta[\chi(t)]$ with the dynamic time-asymmetry of the trajectory.

We note here that the reasoning leading to Eq.(21) is not new. In the absence of delay, this has been used to get the crucial and important relationship between the heat dissipation along a trajectory $q[\chi(t)]$ and the dynamic irreversibility [5, 29], i.e.,

$$\Delta s_m[\chi(t)] = q[\chi(t)]/T = \int_{t_0}^{t_1} F(x, x_{\tau=0}; \lambda)dx = \ln\frac{p(\chi(t)|x_0)}{\tilde{p}(\tilde{\chi}(t)|\tilde{x}_0)}. \quad (22)$$

What we want to emphasize here is that the dynamic irreversibility in a delayed stochastic system, Eq.(22) no longer holds, and the right hand side of Eqs.(21) is not directly related to the thermodynamic variable $q[\chi(t)]$ or $\Delta s_m[\chi(t)]$ anymore, but to a more general dissipation functional $\eta[\chi(t)]$.

A direct consequence of Eq.(21) is that the so-called integral FT holds for the total dissipation $R = \Delta s + \eta$, say, $\langle e^{-R} \rangle = 1$ following the simple equality [4,5]

$$\begin{aligned}\langle e^{-R} \rangle &= \sum_{\chi(t)} p(x_0) p(x(t)|x_0) \frac{p(\tilde{x}_0) p(\tilde{\chi}(t)|\tilde{x}_0)}{p(x_0) p(\chi(t)|x_0)} \\ &= \sum_{\chi(t)} p(\tilde{x}_0) p(\tilde{\chi}(t)|\tilde{x}_0) = \sum_{\tilde{\chi}} p(\tilde{x}_0) p(\tilde{\chi}(t)|\tilde{x}_0) = 1\end{aligned} \quad (23)$$

When delay is absent, this recovers $\langle e^{-\Delta s_{tot}} \rangle = 1$ as shown in Seifert's work [4]. Furthermore, if the initial and final distributions are both chosen from a stationary one, the probability distribution obeys a stronger detailed FT [5]

$$p(R)/p(-R) = e^R. \quad (24)$$

## 3. APPLICATION TO A DELAYED LINEAR SYSTEM

We now consider a delayed linear system subject to periodic external force

$$\dot{x} = ax + bx_\tau + f(t) + \xi(t), \quad (25)$$

where $a < 0$, $|b| < |a|$ and the external force is $f(t) = A_0 \cos(2\pi\omega t)$. The system described by Eq.(25) may be considered as the simplest system with feedback control and periodical driving force, where the effects of inertia are neglected in an overdamping limit. Standard procedures for stochastic dynamics [31] are used to simulate Eq.(25) with a time step $\Delta t = 0.01$. The parameters $a = -0.2$, $D = 0.01$, $\omega = 0.008$ are fixed throughout the present work, while $b, \tau, A_0$ are variable.

We first demonstrate that ensemble average of the total entropy change along a stochastic path, $\langle \Delta s_{tot} \rangle$, is not positive-definite. To this end, we collect $2 \times 10^6$ random trajectories after the system have reached the stationary state, each with time length $t_1 - t_0 = 5$. $\Delta s_m$ is calculated by using Eqs.(7) and (18) for each trajectory. The ensemble of these trajectories is also used to calculate the stationary distribution $p(x(t), t)$, which is required to calculate the system entropy change $\Delta s$. In Fig. 1, the dependences of $\langle \Delta s_{tot} \rangle$ on parameter $b$ are shown for different $\tau$ and $A_0$. Several points can be addressed. When the external force is absent or very small, see Fig.(1a) and (1b), $\langle \Delta s_{tot} \rangle$ could be negative for $b > 0$. For $b < 0$, however, $\langle \Delta s_{tot} \rangle$ is always positive. If $A_0$ is large, $\langle \Delta s_{tot} \rangle$ is dominated by the external force, and $\langle \Delta s_{tot} \rangle$ is also positive. For small (resp. large) $A_0$, $\langle \Delta s_{tot} \rangle$ is a monotonically decreasing (resp. increasing) function of $b$, while for a moderate external force, as shown in Fig.(1c), $\langle \Delta s_{tot} \rangle$ shows a minimum with the change of $b$. Fig.1 also implies that $\langle \Delta s_{tot} \rangle$ does not change much with the delay time $\tau$. To take a closer look, we depict in Fig.2 $\langle \Delta s_{tot} \rangle$ as functions of $\tau$ for different $b$ and $A_0$. The main observation is that $\langle \Delta s_{tot} \rangle$ grows (decreases) slightly for $b > 0$ ($b < 0$), respectively. Since $\langle \Delta s \rangle$ is usually small compared to $\langle \Delta s_m \rangle$, $\langle \Delta s_{tot} \rangle$ is mainly contributed by $\langle \Delta s_m \rangle$, which is associated with the heat dissipation into the reservoir. The interesting dependences of $\langle \Delta s_{tot} \rangle$ on $b$ and $\tau$ thus unravel the connection between the heat dissipation and the feedback control.

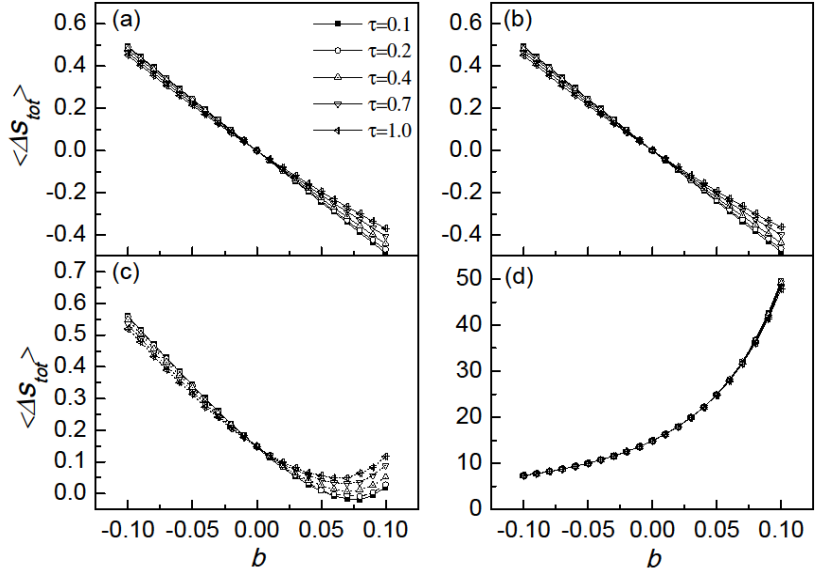

**Figure 1**: The ensemble-averaged total entropy change $\langle \Delta s_{tot} \rangle$ as a function of delay-feedback coefficient $b$. From (a) to (d), the amplitude of external force is $A_0 = 0.0, 0.01, 0.1$ and $1.0$, respectively.

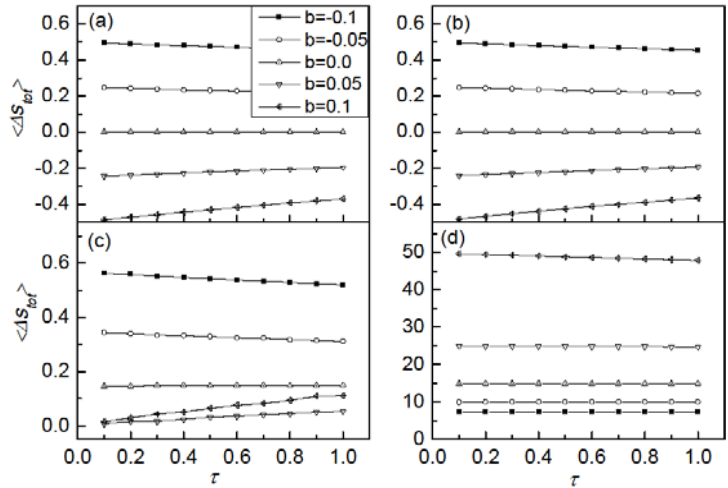

**Figure 2**: The total entropy change $\langle \Delta s_{tot} \rangle$ as a function of delay time $\tau$.

From (a) to (d), $A_0 = 0.0, 0.01, 0.1$ and $1.0$. Note that for $b = 0$, delay does not make sense thus $\langle \Delta s_{tot} \rangle$ does not change with $\tau$.

We note that in Ref.[27], an analytical expression for $\langle \Delta s_{tot} \rangle$ was obtained by using a so-called response function approach. As shown there, $\langle \Delta s_{tot} \rangle$ can be split into two parts, one dependent on the external force and the other not. In the case both $a$ and $b$ are negative, the authors proved that $\langle \Delta s_{tot} \rangle$ is positive definite. Hence, they stated that $\langle \Delta s_{tot} \rangle > 0$ could be considered as the second law of thermodynamics of such a delayed stochastic system. However, as we show here, $\langle \Delta s_{tot} \rangle$ can be negative for $b > 0$, which can also be concluded by taking a closer look into the proof in Ref.[27]. As we discussed in Section 2, one must properly account for the entropy flux induced by the delay feedback to recover the second law, Eq.(16). We will demonstrate this in the following parts.

To check the validity of the second law $\langle R \rangle \geq 0$ and FT $\langle e^{-R} \rangle = 1$ or $p(R)/p(-R) = e^R$, one must first determine the DAF $\bar{F}(x,t)$, which is, however, a rather nontrivial task. Here we adopt the inversion method as that used in Ref.[32]. As mentioned in Ref.[27], when $t \to \infty$, the system will approach a stationary state with distribution $p_{st}(x,t)$. The term 'stationary state' does not mean that $p_{st}(x,t)$ is time-independent, but that the system is in a state attained after relaxation of all transient processes. When $A_0 = 0$, $p_{st}(x,t)$ is time-independent, but when $A_0 \neq 0$, $p_{st}(x,t)$ becomes time-dependent with a time-dependent mean value $x_m(t) = \langle x(t) \rangle_{st}$. As demonstrated in Ref.[27], for the system considered here, the variance $\sigma = \langle x^2(t) - \langle x(t) \rangle^2 \rangle_{st}$ of $x(t)$ is the same whether $A_0$ is zero or not. By introducing $y = x - x_m(t)$, the FPE associated with the distribution of $y$ reads from Eq.(13) as $\partial_t p(y,t) = -\partial_y [\bar{F}(y,t) p(y,t)] + D\partial_y^2 p(y,t)$, which assumes a stationary distribution $p_{st}(y) = N \exp\left((1/D)\int^y \bar{F}(y')dy'\right)$. One can then invert $p_{st}(y)$ to get $\bar{F}(y) = Dd(\ln p_{st}(y))/dy$ [32], from which we can obtain the effective force $\bar{F}(x,t)$. As shown in Ref. [27], [28], and [32], the stationary distribution for $y$ is simply Gaussian,

$$p_{st}(y(t)) = N \exp\left(-\frac{y^2(t)}{2\sigma}\right) \quad (26)$$

where $N$ is a normalization constant. Therefore, the delay averaged force $\bar{F}(x,t)$ readily reads

$$\bar{F}(x,t) = -\frac{D}{\sigma}[x(t) - x_m(t)]. \quad (27)$$

Here, we note that the explicit expressions of the variance $\sigma$ and mean $x_m(t)$ are complicated in general [27-28, 32]. Since the main motivation for us is to check the validity of the second law and FT, we have mainly focused on the case when $\tau$ is small such that much simpler analytical expressions of $\sigma$ and $x_m(t)$ can be obtained which facilitates the numerical calculations. By using a Taylor expansion in powers of $\tau$, we can easily get the expressions for $\sigma$ and $x_m(t)$ as

$$\sigma \cong \frac{(1-b\tau)}{a+b}D, \quad x_m = \frac{A_0(1-b\tau)}{\sqrt{4\pi^2\omega^2 + r^2}}\cos(2\pi\omega t - \theta_0) \quad (28)$$

where $r = (1-b\tau)(a+b)$ and $\theta_0 = \arccos(r/\sqrt{4\pi^2\omega^2 + r^2})$. Eqs. (27) and (28) are thus used for numerical calculations in the following parts of the present work.

From above equations, we are ready to calculate the functional $R[\chi(t)]$ as well as its distribution. To this end, we also collect $2\times 10^6$ random trajectories of length $t_1 - t_0 = 5$ after the system has reached the stationary state. The initial states of the paths are chosen randomly from the stationary distribution. The paths are generated by numerical integration of the original Langevin equation (1). For each trajectory, the integral $\int_{t_0}^{t_1} \bar{F}(x,t)\dot{x}dt$ gives the functional $\eta[\chi(t)]$, wherein $\bar{F}(x,t)$ is obtained by using Eqs.(27) and (28). The system entropy change along the path, $\Delta s[\chi(t)]$, can be readily obtained from the stationary distribution $p_{st}(x_0, t_0)$ and $p_{st}(x_1, t_1)$. The summation of $\eta[\chi(t)]$ and $\Delta s[\chi(t)]$ gives $R[\chi(t)]$, which is used to calculate the ensemble average $\langle R \rangle$ and the distribution $p(R)$.

In Fig.3, $\langle R \rangle$ is plotted as a function of $b$ with or without the external force when $\tau = 0.1$. Clearly, the second law $\langle R \rangle \geq 0$ holds for all values of $b$, in distinct contrast to Fig.1. When $A_0 = 0$, it is shown that $\langle R \rangle \cong 0$, which means that the entropy flux induced by the delay exactly balance the total entropy change generated by the system. In this case, the effective FPE, Eq.(13), satisfies 'detail balance' and the stationary distribution $p_{st}(x)$ can be viewed as the 'equilibrium' Boltzman distribution of an overdamped Brownian particle in an effective time-independent harmonic potential $V_{eff}(x) = \frac{k_B T}{2\sigma} x^2$. When $A_0 \neq 0$, detail balance of the FPE is broken and we find that $\langle R \rangle > 0$ due to non-equilibrium resulting from the external force.

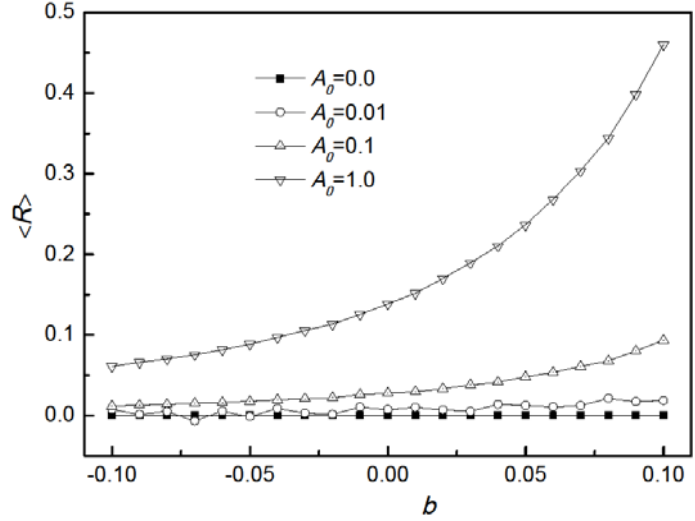

**Figure 3**: The total dissipation functional $\langle R \rangle$ as a function of the delay-feedback coefficient $b$. The second law holds in the form $\langle R \rangle \geq 0$. Note that the data shown for $A_0 = 0.01$ have been multiplied by 20, and that for $A_0 = 1.0$ have been divided by 100.

We have also verified the validity of the integral or detailed FT for $R$. When $A_0 = 0$, the distribution of $R$ is a $\delta$-like function around $R = 0$, and $\langle e^{-R} \rangle = 1$ holds to a very good accuracy, as shown in Fig.4. For comparison, the distribution of $\Delta s_{tot}$ and the value of $\langle e^{-\Delta s_{tot}} \rangle$ are also shown in Fig.4. Clearly, the distribution of $\Delta s_{tot}$ is much wider and the discrepancy between $\langle e^{-\Delta s_{tot}} \rangle$ and 1 is quite apparent. Also note that for positive feedback, e.g., $b = 0.1$, $\Delta s_{tot}$ peaks around a negative value as shown in Fig.(4a), such that $\langle e^{-\Delta s_{tot}} \rangle > 1$ as shown in Fig.(4b). When $A_0 \neq 0$, typical distributions of $R$ are shown in Fig.(5a), which are not symmetric around zero and not Gaussian. Obviously, trajectories with negative $R$ contribute considerably to the distribution, which are associated with second-law-violation events. Since direct verification of the integral FT usually requires extensive statistics of these negative-$R$ events, we turn to check the detailed FT, Eq.(24), which is stronger than the integral FT but easier for numerical demonstration. As shown in Fig.(5b), $p(R)/p(-R) = e^R$ does hold to a good accuracy. In contrast, the distribution of $\Delta s_{tot}$, also presented in Fig.(5a), does not satisfy this FT, see also Fig.(5b).

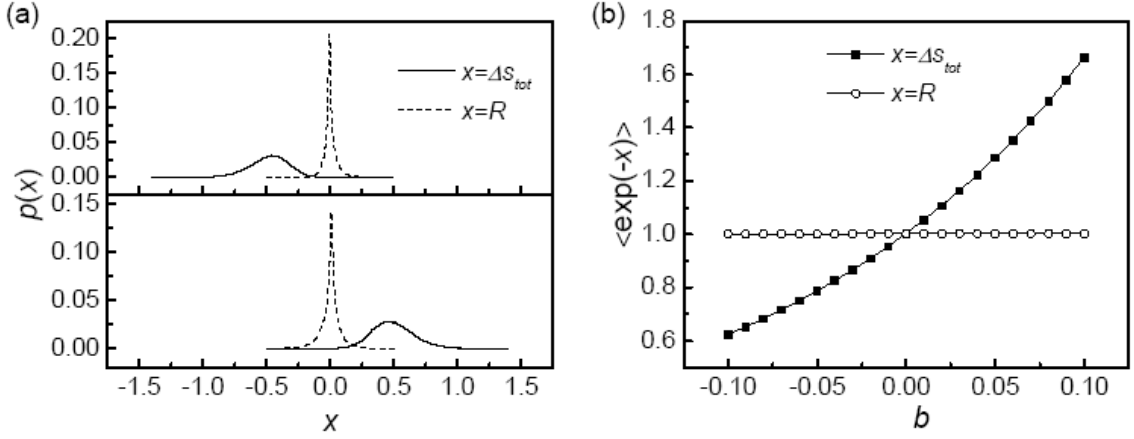

**Figure 4:** Validation of the integral FT when $A_0 = 0$. (a) Distributions of $\Delta s_{tot}$ and $R$ for $b=0.1$ (top) and $b=-0.1$ (bottom). (b) $\langle \exp(-x) \rangle$ as a function of the coefficient $b$, where $x$ denotes $\Delta s_{tot}$ (squares) or $R$ (circles). The integral FT $\langle e^{-x} \rangle = 1$ holds true for $R$ but not for $\Delta s_{tot}$.

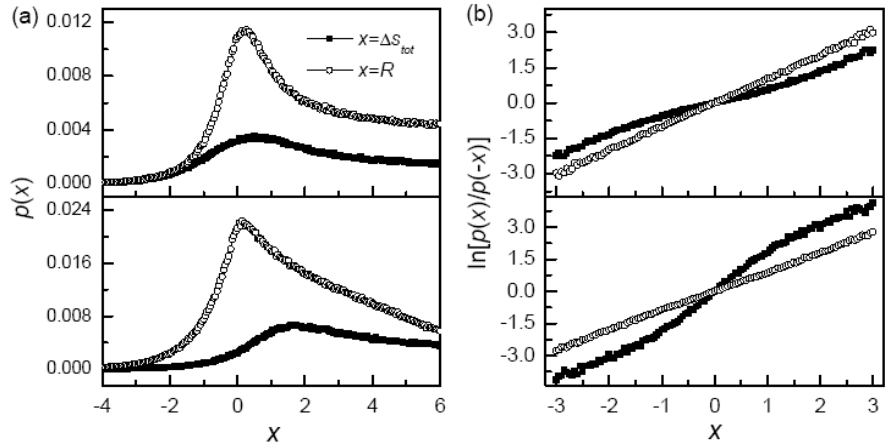

**Figure 5:** Validation of the detailed FT for $R$ whe $A_0 = 1.0$. (a) Distributions of $\Delta s_{tot}$ or $R$ for $b=0.1$ (top) and $b=-0.1$ (bottom). (b) $\ln\left[p(x)/p(-x)\right]$ as a function of $x$ for $b=0.1$ (top) and $b=-0.1$ (bottom), where $x$ stands for $\Delta s_{tot}$ or $R$. The detailed FT $p(x)/p(-x) = e^x$ holds true for $R$ but not for $\Delta s_{tot}$.

One should note here that direct extension of the above numerical methods to more complex delayed stochastic systems is not straightforward. The main reason is that it is hard, in general, to obtain the stationary distribution $p_{st}(x,t)$, which arises from the fact that analysis of delayed stochastic systems is rather nontrivial, and many open questions remain unsolved. However, the main results of the present paper, i.e., correct interpretations of the second law and FTs, hold true whether numerical demonstrations are feasible or not.

## 4. CONCLUSION

In summary, we have considered the stochastic thermodynamics of a Langevin system with time delay, where the trajectory dependent thermodynamic quantities are introduced to study the first law and second law. The time delay brings new feature to the system, i.e., additional delay-entropy flux is injected to the system during the physic process. The total entropy production $\Delta s_{tot} = \Delta s + \Delta s_m$ is no more a good criterion for the second law. We suggest the inequality $\langle R \rangle \equiv \langle \Delta s + \eta \rangle \geq 0$ as the second law in delayed stochastic systems, where $\eta$ is a delay-averaged trajectory-dependent dissipation functional. With the help of a Fokker-Planck description, we can reconstruct the integral and detailed FT for $R$. In a linear model with time delay, we give numerical evidences that the total entropy production $\langle \Delta s_{tot} \rangle$ could indeed be negative when the delay feedback is positive. By inversing delay-averaged force from the stationary distribution, we have calculated the total dissipation functional $R$ as well its distribution. The second law and FTs are successfully reproduced by simulations. Since stochastic systems with time delay are of ubiquitous importance in realistic nano-systems, we hope that the present study could arise experimental research interests and open more perspectives on the study of nonequilibrium thermodynamics in small systems.

**Acknowledgments**:   The work is supported by the National Science Foundation of China (Grant No. 20873130, 20933006).